\documentclass[prl, twocolumn, superscriptaddress, longbibliography]{revtex4-1}
\usepackage{amsmath}
\usepackage{amssymb}
\usepackage{graphicx}
\usepackage[caption=false, position=top, singlelinecheck=off, justification=raggedright]{subfig}
\usepackage{color}
\usepackage{pst-node}
\usepackage[unicode]{hyperref}
\hypersetup{unicode=true, colorlinks=true, linkcolor=blue, citecolor=blue}

\begin{document}

\title{Evidence for Quantum Stripe Ordering in a Triangular Optical Lattice}
\author{Xiao-Qiong Wang}
\thanks{These authors have contributed equally to this work.}
\affiliation{Department of Physics, Southern University of Science and Technology, Shenzhen 518055, China}
\affiliation{Shenzhen Institute for Quantum Science and Engineering, Southern University of Science and Technology, Shenzhen 518055, China}
\author{Guang-Quan Luo}
\thanks{These authors have contributed equally to this work.}
\affiliation{Department of Physics, Southern University of Science and Technology, Shenzhen 518055, China}
\affiliation{Shenzhen Institute for Quantum Science and Engineering, Southern University of Science and Technology, Shenzhen 518055, China}
\author{Jin-Yu Liu}
\affiliation{Department of Physics, Southern University of Science and Technology, Shenzhen 518055, China}
\affiliation{Shenzhen Institute for Quantum Science and Engineering, Southern University of Science and Technology, Shenzhen 518055, China}
\author{Guan-Hua Huang}
\affiliation{Department of Physics, Southern University of Science and Technology, Shenzhen 518055, China}
\affiliation{Shenzhen Institute for Quantum Science and Engineering, Southern University of Science and Technology, Shenzhen 518055, China}
\author{Zi-Xiang Li}
\affiliation{Institute of Physics, Chinese Academy of Sciences, Beijing 100190, China}
\author{Congjun Wu}
\affiliation{New Cornerstone Science Laboratory, Department of Physics,
School of Science, Westlake University, 310024, Hangzhou, China}
\affiliation{Institute for Theoretical Sciences, Westlake University, 310024, Hangzhou, China}
\affiliation{Key Laboratory for Quantum Materials of Zhejiang Province, Department of Physics,
School of Science, Westlake University, Hangzhou 310030, China}
\affiliation{Institute of Natural Sciences, Westlake Institute for Advanced Study, Hangzhou 310024, China}
\author{Andreas Hemmerich}
\email{hemmerich@physnet.uni-hamburg.de}
\affiliation{Institute of Quantum Physics, University of Hamburg, Luruper Chaussee 149, 22761 Hamburg, Germany}
\author{Zhi-Fang Xu}
\email{xuzf@sustech.edu.cn}
\affiliation{Department of Physics, Southern University of Science and Technology, Shenzhen 518055, China}
\affiliation{Shenzhen Institute for Quantum Science and Engineering, Southern University of Science and Technology, Shenzhen 518055, China}

\begin{abstract}
Understanding strongly correlated quantum materials, such as high $T_\textrm{c}$ superconductors, iron-based superconductors, and twisted bilayer graphene systems, remains to be one of the outstanding challenges in condensed matter physics. Quantum simulation with ultra-cold atoms in particular optical lattices, which provide orbital degrees of freedom, is a powerful tool to contribute new insights to this endeavor. Here, we report the experimental realization of an unconventional Bose-Einstein condensate of $^{87}$Rb atoms populating degenerate $p$-orbitals in a triangular optical lattice, exhibiting remarkably long coherence times. Using time-of-flight spectroscopy, we observe that this state spontaneously breaks the rotational symmetry and its momentum spectrum agrees with the theoretically predicted coexistence of exotic stripe and loop current orders. Like certain strongly correlated electronic systems with intertwined orders, as high-$T_\textrm{c}$ cuprate superconductors, twisted bilayer graphene, and the recently discovered chiral density-wave state in kagome superconductors $\textrm{AV}_3 \textrm{Sb}_5$ (A=K, Rb, Cs), the newly demonstrated quantum state, in spite of its markedly different energy scale and the bosonic quantum statistics, exhibits multiple symmetry breakings at ultralow temperatures. These findings hold the potential to enhance our comprehension of the fundamental physics governing these intricate quantum materials.
\end{abstract}

\maketitle

\begin{figure}
\centering
\includegraphics[width=\linewidth]{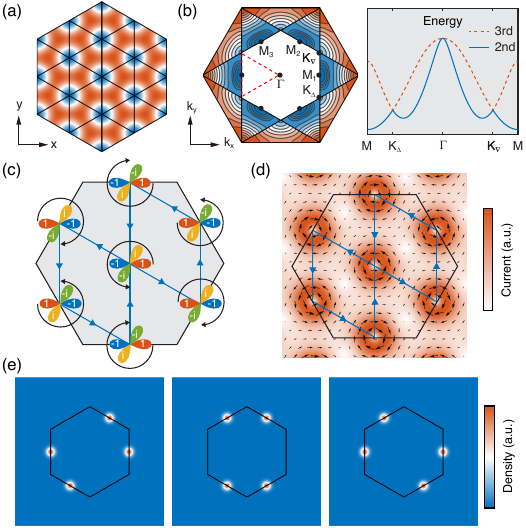}
\caption{Quantum stripe ordering for $p$-orbital bosons in a triangular optical lattice. (a) Lattice potential of the triangular optical lattice with maxima and minima shown by red and blue color, respectively. (b) The left panel shows a contour plot of the second and third Bloch bands across the second and third Brillouin zones, respectively. The energy dispersion along the high symmetry lines forming the dashed red triangle is shown for both bands in the right panel. The three inequivalent $M$-points provide degenerate minima protected by the discrete rotational symmetry of the lattice. (c) Schematic order parameter of one of six possible implementations of the ground states within the second band. The gray hexagon denotes the unit cell of the order parameter. Each site contains an equal hybridization of two non-orthogonal $p$-orbitals with a relative phase difference $\pm\pi/2$. The circular arrows illustrate the on-site orbital angular momentum. The solid blue lines with arrows denote the bond currents between nearest neighbor sites. (d) The corresponding mass current of  the ground state. The amplitude (direction) of the current is shown by a color code (black arrows). (e) Momentum distributions of three representative ground states predicted theoretically. Solid black hexagons denotes the first Brillouin zone.}
\label{fig1}
\end{figure}

Understanding the foundational principles of quantum matter is of central importance in both condensed matter and cold atom physics. A hallmark of phase diagrams of correlated quantum materials is the emergence of multiple intertwined electronic orders~\cite{KivelsonReview, proust2019}. For example, in unconventional superconductors (SCs)~\cite{KeimerReview}, superconductivity is intertwined with multiple orders including antiferromagnetism~\cite{scalapino2012common}, nematicity~\cite{fernandes2014drives}, pair-density wave~\cite{Agterberg2020}, and stripe order~\cite{tranquada1995evidence}, emerging at different temperature scales and doping concentrations. Similar intertwined orders with different broken symmetries are ubiquitous in correlated systems, including iron-based superconductors\cite{Si2016}, twisted bilayer graphene~\cite{Cao2018,Andrei2020}, and kagome superconductors~\cite{Jiang2021, Mielke2022, Nie2022, Zhao2021, Chen2021, jiang2021kagome,Neupert2022}. Unravelling the physics underlying these intertwined orders will lead to a significant leap in understanding the fundamental principles of exotic quantum states in general.

The microscopic mechanisms driving intertwined orders in quantum materials like cuprate and Kagome superconductors are still widely unexplained. Unveiling their fundamental physics requires precise tuning of system parameters and appropriate methods to observe and discriminate different orders. This can be an exceedingly difficult task due to the vast complexities of solid state quantum materials. In contrast, ultracold atom systems offer an unchallenged combination of precision and control with respect to a wide spectrum of physical properties like quantum statistics, lattice geometry, dimensionality or interaction strength~\cite{Bloch2005, Lewenstein2007, Gross2017, Takahashi2020}. Unconventional lattice geometries, e.g., triangular ~\cite{Sengstock2010, Sengstock2011}, honeycomb~\cite{Soltan2011, Jin2021, Wang2021} or kagome~\cite{Jo2012}, have been realized. Techniques as Floquet engineering~\cite{Simonet2021} and the implementation of orbital degrees of freedom~\cite{Wu2009, Li2016, Kock2016} enable investigations beyond conventional s-band physics. Hence, ultracold atom can serve as complementary platform for exploring intertwined orders, including nematicity, loop currents, and stripe order.

Here, we report the first experimental realization of a Bose-Einstein condensate (BEC) of neutral atoms in the $p$-orbitals of a triangular optical lattice. Through active cooling during the dissipative condensation dynamics, we are able to largely extend the coherence time, such that the metastable $p$-orbital~\cite{Wu2009, Kock2016, Li2016} ground state is closely approached. Remarkably, the interplay between lattice frustration and orbital degeneracy gives rise to an exotic finite-momentum superfluid phase reminiscent of pair-density wave states in SCs~\cite{Agterberg2020, Agterberg2011}. It also spontaneously breaks the sixfold lattice rotational symmetry. Unequivocal evidence is observed using time-of-flight spectroscopy.  The observed momentum spectra clearly confirm the quantum stripe phase and are consistent with its coexistence with a loop current order, theoretically predicted for bosons in the $p$-orbitals of the triangular lattice~\cite{Wu2006}. 
Note that loop current order is proposed as a candidate for the pseudogap state in high-$T_\textrm{c}$ cuprate SC~\cite{varma1997}, attracting immense attention in both theoretical~\cite{capponi2004} and experimental regards~\cite{Bourges2021}. Nevertheless, despite a continuing quest during the past two decades, unequivocal experimental evidence for loop currents has not been found. Recently, possible signatures of loop currents were detected in bilayer graphene systems~\cite{Liu2021} and kagome SCs~\cite{Jiang2021, Mielke2022, Nie2022, jiang2021kagome, Neupert2022}. Thus, the experimentally realized quantum stripe phase is expected to capture aspects of a bosonic version of the intertwined orders in cuprate high-$T_\textrm{c}$ SC and kagome SC $\textrm{AV}_3 \textrm{Sb}_5$ (A=K, Rb, Cs), which could provide new insights into the fundamental physics of these correlated quantum materials. 
Moreover, our findings show remarkable differences as compared to the chiral superfluidity obtained in square~\cite{Wirth2011} and hexagonal bipartite optical lattices~\cite{Wang2021}, where the lattice rotational symmetry is preserved and no bond currents arise between adjacent sites. This work also differs significantly from previous works where time-reversal symmetry is broken at a single-particle level via applying a synthetic magnetic field~\cite{Aidelsburger2013, Miyake2013,  Aidelsburger2014, Kennedy2015} and hence it is not a surprise to obtain a nonzero circulating current~\cite{Atala2014}. The unconventional multi-orbital superfluidity discussed in Ref.~\cite{Soltan_Panahi2011} with three-fold rotational symmetry is not a near-equilibrium state of the system but rather a dynamical artifact. The symmetry breaking from $C_6$ to $C_3$ is due to the interaction during the time-of-flight expansion dynamics~\cite{Weinberg2016}.

\begin{figure*}[!t]
\centering
\includegraphics[width=\linewidth]{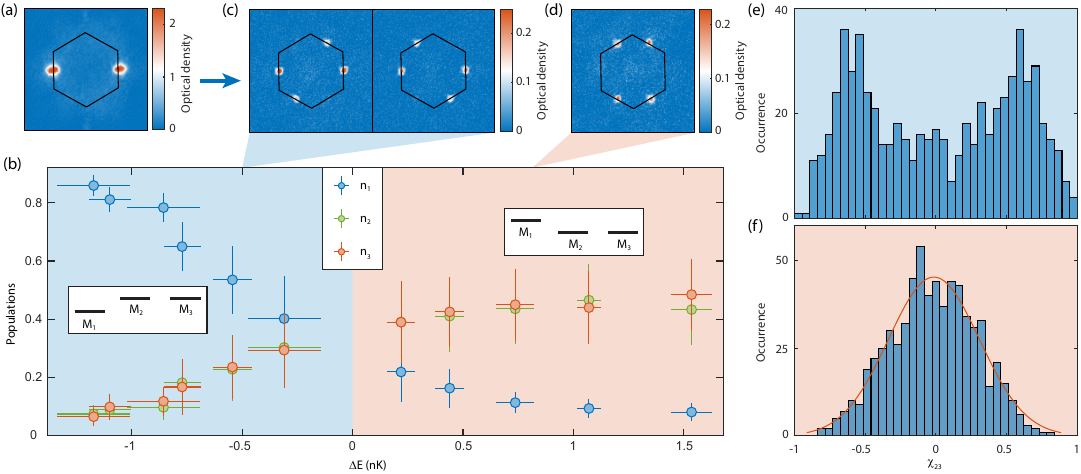}
\caption{Quantum phase transition induced by fine tuning the energy imbalance.  (a) The momentum distribution for the initial state, $0.1\,$ms after preparation of a condensate at $M_1$. (b) A distortion of the triangular lattice is applied to adjust the energy imbalance $\Delta E\equiv E_2(M_1)-[E_2(M_2)+E_2(M_3)]/2$ with equal energies $E_2(M_2) = E_2(M_3)$ maintained. The $M$-point populations after $130.1\,$ms evolution time, averaged over 111 experimental runs, is plotted versus $\Delta E$. Error bars denote standard deviations. The insets illustrate the two characteristic configurations with respect to the $M-$point energies for $\Delta E < 0$ and $\Delta E > 0$. (c) Exemplary final state momentum distributions, recorded by time-of-flight measurements with $\Delta E\approx -0.5\, \rm nK$ as indicated in the figure. (d) Same as (c), but with $\Delta E \approx 0.8\, \rm nK$. (e) and (f) Histograms for the occurrence frequencies of the single-run values of $\chi_{23}$, recorded after 140.1 ms holding time, for more than 600 runs with the same values of $\Delta E$ used in (c) and (d), respectively. The red solid line in (f) shows a Gaussian fit.}
\label{fig2}
\end{figure*}

Our experiments use a BEC of rubidium atoms. The atoms are optically confined by a dipole trap and a two-dimensional triangular optical lattice. The dipole trap is created by adding potentials generated respectively from three non-interfering laser beams propagating within the $xy$-plane, mutually intersecting at an angle of $120^{\circ}$. The lattice potential (shown in Fig.~\ref{fig1}(a)) is formed by three interfering laser beams, polarized linearly along the $z$-axis, and propagating in opposite directions with regard to the three dipole-trap laser beams. All beams have a wavelength $\lambda$ of about $1,064\, \rm nm$. This specific configuration ensures that the combined potential of the triangular lattice and the dipole trap practically maintains a sixfold rotational symmetry. Focusing on the $p$-orbital Bloch bands of the lattice (i.e., the second and third bands), a triple-well scenario in the quasi-momentum space arises in the second band, i.e., one finds three inequivalent energy minima of the second band, denoted $M_1, M_2, M_3$ in Fig.~\ref{fig1}(b), located at the centers of the edges enclosing the first Brillouin zone. In our experiment we can adjust the second band energies $E_2(M_i), i\in \{1,2,3\}$ at the $M$-points via distortions of the unit cell (cf. Supplementary Materials). For the maximally symmetric case $E_2(M_1)=E_2(M_2)=E_2(M_3)$, theoretical considerations predict that the interaction among $p$-orbital bosons favors a metastable BEC~\cite{Wu2006}, which equally populates two of three $M$-points, and thus simultaneously breaks time-reversal symmetry (TRS) and the sixfold lattice rotational symmetry accompanied by nonzero staggered loop currents. One of three possible implementations of the ground state in the second Bloch band is sketched in Fig.~\ref{fig1}(c), which shows two $p$-orbitals at each lattice site enclosing an angle of $60^ \circ$ superimposed with relative phases $\pm\pi/2$ leading to nonzero bond currents between nearest neighbor sites. The associated antiferromagnetic angular momentum is illustrated in Fig.~\ref{fig1}(d). In Fig.~\ref{fig1}(e), predicted momentum spectra of the three possible ground states are shown, each characterized by population peaks at different combinations of two $M$-points. 

We start with a bipartite hexagonal lattice, composed of two triangular lattices, slightly deformed such that the second band possesses a global minimum at a single $M$-point (e.g., $M_1$). Details are found in Ref.~\cite{suppl}. This allows us to load a BEC into the second band at that $M$-point by rapidly switching the relative potential offset $\Delta V$ between the two available classes of potential wells (cf. Ref.~\cite{Wang2021}, Methods and Supplementary Information). By choosing the final value of $\Delta V$ appropriately, we ensure that, in configuration space, predominantly the $s$-orbitals of the shallow wells are occupied. As discussed in Ref.~\cite{Wirth2011,Kock2016}, this keeps band relaxation collisions at a low level. In a second, adiabatic step, one of the triangular lattices is completely turned off such that the atoms in configuration space are transferred to the $p$-orbitals of the other triangular lattice, while in regard to momentum space, they remain localized at the chosen $M_1$-point. Once the $p$-orbitals are populated, increased band relaxation and associated heating sets in. After preparation of the BEC at the $M_1$-point shown in Fig.~\ref{fig2}(a), the system is allowed to relax to its ground state. In order to maintain the phase coherence necessary to form the theoretically predicted quantum stripe ordering~\cite{Wu2006}, we continuously cool the system by evaporation at the cost of extra atom loss.

We first simplify the symmetry breaking process necessary to reach one of degenerate ground states in the limited phase coherence time. Therefore, the degeneracy of the $M$-points is partly lifted by inducing a small power imbalance of the three laser beams forming the triangular lattice. We fine tune the energy imbalance $\Delta E\equiv E_2(M_1)-[E_2(M_2)+E_2(M_3)]/2$ while maintaining the degeneracy of $M_2$ and $M_3$ with $E_2(M_2)=E_2(M_3)$. Fig.~\ref{fig2}(b) records the averaged condensate populations $n_j$ at $M_j$ with $j=1,2,3$ and $\sum_j n_j=1$ after 130.1 ms holding time obtained from momentum distributions recorded via time-of-flight measurements. Starting from large and negative $\Delta E$, interactions then favor almost all atoms condensing at $M_1$ with tiny populations at the other two $M$ points. With increasing $\Delta E$, the average population for atoms at $M_1$ monotonically decreases and the populations at $M_2$ and $M_3$ increase accordingly with equal average values. Characteristic momentum distributions for negative and positive values of $\Delta E$ are shown in Fig.~\ref{fig2}(c) and Fig.~\ref{fig2}(d). 

When $\Delta E<0$, we find that the atoms prefer to condense at two out of three $M$ points with the majority of the atoms populating the $M_1$ point with lower energy, and the minority of the atoms residing at the higher energy of one of the two degenerate $M_2$ or $M_3$ points. This is consistent with the histogram shown in Fig.~\ref{fig2}(e), where a double peak structure appears for the population imbalance given by $\chi_{23}\equiv (n_2-n_3)/(n_2+n_3)$ and a negative correlation among atoms populating the $M_2$ and $M_3$ points arises. This indicates that a spontaneous symmetry breaking emerges due to interactions among $p$-orbital atoms. For the other case with $\Delta E>0$, a different scenario occurs. We observe that the majority of atoms prefers to condense at the two lower energy points $M_2$ and $M_3$ with almost equal populations in each single run. This corresponds to the histogram for $\chi_{23}$ shown in Fig.~\ref{fig2}(f), where a Gaussian profile with a maximum value at zero is obtained and a positive correlation among atoms residing at $M_2$ and $M_3$ is found. Based on these experimental results with moderate $|\Delta E|$, the interactions are found to favor the majority of the atoms to condense at two out of three $M$ points. We thus expect that in the limit of $\Delta E\rightarrow 0$, this feature should persist. Theoretically, a first order quantum phase transition occurs when we cross $\Delta E=0$. Details can be found in Ref.~\cite{suppl}. The main feature shown in Fig.~\ref{fig2} is consistent with this prediction. At zero temperature the populations are expected to undergo a discontinuity. Due to the finite temperature, in the experiment this jump is replaced by a continuous transition within a small region of $|\Delta E|$ of a few nanokelvin.  

\begin{figure}
\centering
\includegraphics[width=\linewidth]{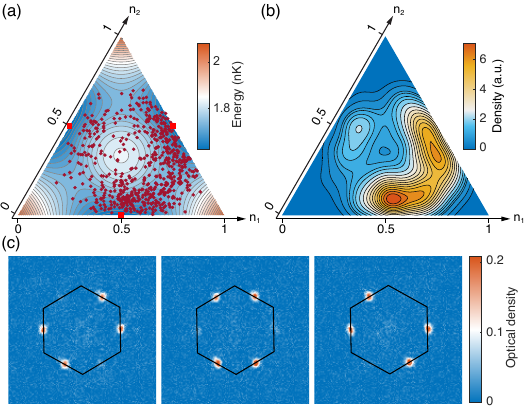}
\caption{Statistical analysis of the quantum stripe phase. (a) Statistical distribution of the final states at $140.1\,$ms holding time for 636 experimental runs. Each dark red dot shows the tuple $(n_1, n_2)$ of populations of the points $M_1$ and $M_2$ using an oblique coordinate system. The mean-field energy per atom calculated via choosing optimized relative phases between condensates at three $M$ points is shown by the color code and contour lines. Three red squares indicate the locations of the theoretically predicted ground states. The corresponding energy scale for the tunneling between two nearest neighbor $p$-orbitals is about $8\,\rm nK$. (b) The kernel density estimation of the data shown in (a), where three local maxima appear close to the predicted ground states, respectively. (c) Momentum distribution for the final states observed close to the three local maxima in the kernel density distribution shown in (b).}
\label{fig3}
\end{figure}

We next consider the maximally symmetric case with three degenerate minima of the second band at the three $M$-points. Theoretical calculations suggest that the atoms should condense at two out of three degenerate Bloch states $\Phi_{j}$ with $\pm \pi/2$ phase difference, selected in a spontaneous symmetry breaking process. The plots in Fig.~\ref{fig1}(e) shows the theoretically predicted atomic momentum distributions for different ground states. To explore the predicted quantum stripe order, we carefully adjust the energy imbalance $\Delta E$ by changing the laser power of three beams and applying parametric heating to calibrate its value. However, this method has limitations with regard to the accuracy. Especially for the balanced case, tiny differences are hard to be precisely calibrated. Nevertheless, we roughly estimated $ |\Delta E|\le 0.3\,\rm nK$. We thus record the momentum distributions for atoms relaxing in the $p$-orbitals for various holding times. Our experiments show two different time scales. First, in a relatively short time interval of about $40\,$ms, the system quickly thermalizes to states with approximately equal populations at three $M$ points. Subsequently, slow evaporation dynamics further cools the system and drives it into lower energy states. These are shown in Ref.~\cite{suppl}. Fig.~\ref{fig3}(a) shows the statistical distribution of the final states after 140.1~ms holding time. Each dark red dot represents one of 636 experimental runs, specifying the populations $n_1$, $n_2$ of the points $M_1$ and $M_2$ using an oblique coordinate system. The mean-field energy per atom, calculated via choosing optimized relative phases between condensates at three $M$ points, is shown by the color code and contour lines~\cite{suppl}. Three red squares indicate the locations of the theoretically predicted ground states.  We apply a non-parametric kernel density estimation method to the data in Fig.~\ref{fig3}(a), using the intrinsic function of MATLAB. The corresponding density distribution is shown in Fig.~\ref{fig3}(b). {\it A deep local minimum emerges close to the center of the triangle, which indicates that interaction does not favor equal populations of all three $M$ points leading to lattice rotational symmetry breaking.} However, three local maxima for the distribution are found to be close to the middle points of the three edges of the triangle (see red squares in Fig.~\ref{fig3}(a)), which are the locations of the theoretical predicted ground states. Corresponding to the three maxima in the distribution in Fig.~\ref{fig3}(b), three exemplary momentum distributions for the final states are shown in Fig.~\ref{fig3}(c), which well reproduce the theoretical predicted momentum spectra of the ground states shown in Fig.~\ref{fig1}(e). These findings provide clear experimental evidence for the emergence of an unconventional BEC of $p$-orbital bosons in a triangular optical lattice, where two out of three $M$ points are populated. Hence, we unambiguously show that three-fold rotational symmetry is spontaneously broken in our experiment. Meanwhile, a quantum stripe order is formed, since atoms condense at two different quasi-momenta.

To determine whether the observed quantum stripe phase also breaks TRS depends upon information of the relative phase of the condensate fractions at the two $M$ points. This typically requires the observation of interference, for example, similar to what has been reported for a bipartite square lattice in Ref.~\cite{Kock2015}. In the triangular lattice of this work, where persistent evaporation cooling is necessary to maintain a low temperature and phase coherence, such measurements are not easily possible. Nevertheless, we can partially infer the desired phase information from a simple mean field consideration. It is reasonable to assume that after a long relaxation time the final state should have a lower interaction energy per particle than the initial state, where all atoms are condensed at the $M_1$ point. Our observations combined with mean field calculations let us exclude the possibility that the phase difference between the condensates at the two $M$ points is close to zero or $\pi$ with the consequence of nonzero bond currents~\cite{suppl}. This gives a clear indication that the observed final states break TRS and show a loop-current order. However, a conclusion merely based upon observations require extensive additional experimental efforts.

To conclude, we have obtained unequivocal evidence for the spontaneous rotational symmetry breaking of an unconventional BEC in the $p$-orbitals of a triangular optical lattice. The observed momentum spectra, according to which the atoms predominantly choose to condense in two of three possible $M$-points, agree with the theoretically predicted coexistence of a quantum stripe phase with a loop current order. Observations combined with mean-field considerations indicate that the observed phase breaks TRS. In summary, the realized exotic phase implements a bosonic analog to the intertwined orders found in certain electronic condensed matter systems.  The ability to mimic intertwined orders in a well-controlled environment offers an opportunity to deepen our understanding of the underlying physics in correlated quantum materials. The achieved long phase coherence in a pure $p$-orbital band could also pave the way to achieve the preparation of strongly correlated states with orbital degrees of freedom~\cite{Tokura2000}, for example, unconventional orbital Mott phases~\cite{Li2016}. 

\begin{acknowledgments}
This work is supported by the National Key R\&D Program of China (Grants No.~2022YFA1404103 and No.~2018YFA0307200), the Key-Area Research and Development Program of Guangdong Province (Grant No.~2019B030330001),  NSFC (Grant No.~12274196 and No.~U1801661), and funds from Guangdong province (Grant No.~2019QN01X087 and Grant No.~2019ZT08X324). C.W. is supported by NSFC (Grant No.~12234016 and No.~12174317) and the New Cornerstone Science Foundation. A.H. acknowledges support by Cluster of Excellence CUI: Advanced Imaging of Matter of the Deutsche Forschungsgemeinschaft (DFG) - EXC 2056 - project ID 390715994.
\end{acknowledgments}

\bibliography{main_text_refs}

\newpage
\onecolumngrid
\renewcommand\thefigure{S\arabic{figure}}
\setcounter{figure}{0}
\renewcommand\theequation{S\arabic{equation}}
\setcounter{equation}{0}
\makeatletter
\newcommand{\rmnum}[1]{\romannumeral #1}
\newcommand{\Rmnum}[1]{\expandafter\@slowromancap\romannumeral #1@}
\makeatother

\newpage

{
	\center \bf \large
	Supplemental Material for: \\
	Evidence for Quantum Stripe Ordering in a Triangular Optical Lattice
	\vspace*{0.1cm}\\
	\vspace*{0.0cm}
}

\vspace{4ex}

\section{\bf Triangular optical lattice}
The two-dimensional triangular optical lattice is realized by superimposing three running-wave laser beams with out-of-plane polarizations that mutually intersect in the $xy$-plane at an angle of $120^{\circ}$. The beams are derived from the same laser emitting at a wavelength $\lambda \simeq 1064$ nm. This results in the two-dimensional lattice potential 
\begin{eqnarray}
V(\mathbf{r})=-V_A\left[ 3+2\sum_{j=1}^3\alpha_{j}\cos \left( \mathbf{b}_j\cdot \mathbf{r} \right) \right]\, ,
\label{potential}
\end{eqnarray}
where $\boldsymbol{\alpha}=(\alpha_{1}, \alpha_{2}, \alpha_{3})$ accounts for small differences in the intensities of the three superimposed beams, which can be controlled by fine tuning their powers. The vectors $\mathbf{b}_j$ are defined as $\mathbf{b}_1=\mathbf{k}_1-\mathbf{k}_2$, $\mathbf{b}_2=\mathbf{k}_2-\mathbf{k}_3$, and $\mathbf{b}_3=\mathbf{k}_3-\mathbf{k}_1$ with the wave vectors ${\bf k}_1=k_L(-\sqrt{3}/2,1/2)$, ${\bf k}_2=k_L(\sqrt{3}/2,1/2)$, and ${\bf k}_3=k_L(0,-1)$, where $k_L=2\pi/\lambda$ and $j=1,2,3$. In order to largely maintain the lattice rotational symmetry, the optical dipole trap is formed by three crossed laser beams propagating within the $xy$-plane along opposite directions with respect to the three lattice beams.

\section{\bf State preparation and detection}
We employ a second auxiliary triangular lattice, which allows us to transfer atoms from its lowest $s$-band to the $p$-band of the other triangular lattice. Both lattices are formed by the same laser beams discussed in the context of Eq.~(\ref{potential}), by providing two emission frequencies $\nu_1$ and $\nu_2$ for each beam with a frequency difference of a few GHz. The total lattice potential is given by
\begin{eqnarray}
V(\mathbf{r})=-V_A\left[3+2\sum_{j}\cos \left( \mathbf{b}_j\cdot \mathbf{r}\right)  \right]  
-V_B\left[3+2\sum_{j}\cos \left( \mathbf{b}_j\cdot \mathbf{r} -\Delta\eta_j\right)  \right].
\label{bn}
\end{eqnarray}
Here, $\Delta\eta_{j}=2\pi\Delta\nu \Delta L_j/c$ determine the lattice center difference between the two triangular lattices. Experimentally, we choose $(\Delta L_1,\Delta L_2,\Delta L_3)=(-6.04,3.02,3.02) \, \rm cm$, which can be tuned by changing the optical paths of the laser beams forming the triangular lattices. The frequency difference $\Delta\nu=\nu_1-\nu_2$ can be precisely adjusted experimentally. If $\Delta\nu=3.308\,$GHz, $V(\mathbf{r})$ is a hexagonal boron nitride lattice, while if $\Delta\nu=3.25\,$GHz, it is a deformed hexagonal lattice. The energy minimum of the second band moves from the $K$ points to the $M_1$ point, when $\Delta\nu$ is tuned from 3.308~GHz to 3.25~GHz.

\begin{figure}[htbp]
	\centering
	\includegraphics[width=12cm]{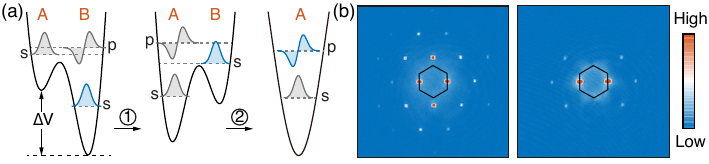}
	\caption{(a) The experimental sequence for transferring atoms from the $s$-orbital of a $B$-site to the $p$-orbital of an $A$-site. The lattice depths $(V_A, V_B)$ from the left to the right are $(6.96, 7.92)~E_{\rm{R}}$, $(7.82, 7.20)~E_{\rm{R}}$ and $(0.71, 0)~E_{\rm{R}}$, respectively. (b) The left image shows a the momentum spectrum after step 1 of (a) followed by 165~ms holding time, while the right image shows a momentum spectra after step 2 of (a) followed by 0.1~ms holding time.}
	\label{sfig1}
\end{figure}

Our experiment starts with a $^{87}$Rb Bose-Einstein condensate in the $|F=1,m_F=-1\rangle$ state, with $\sim 1.3\times10^5$ atoms confined in a crossed optical dipole trap with the trapping frequencies $\lbrace \omega_x,\omega_y,\omega_z \rbrace=2 \pi \times \lbrace 26.4, 26.5, 69.5\rbrace\,$Hz. The atoms are then loaded into the deformed hexagonal lattice arising for $\Delta\nu=3.25\,$GHz with $(V_A, V_B)=(6.96, 7.92)~E_{\rm{R}}$, with the recoil energy $E_\mathrm{R}=h^2/2m\lambda^2$. By further tuning to $(V_A, V_B)=(7.82, 7.20)~E_{\rm{R}}$, the $M_1$ point of the deformed hexagonal lattice is populated. Finally, the atoms are transferred into the $p$-band of the $A$ triangular lattice by linearly ramping $(V_A, V_B)$ to $(0.71, 0)~E_{\rm{R}}$ in 11.5~ms. The experimental sequence and the results are shown in Fig.~\ref{sfig1}. As the lattice depth is decreased, the trapping potential provided by the superposition of the lattice and the optical dipole trap is insufficient to trap atoms along the direction of gravity. Therefore, we linearly increase the intensity of the optical dipole trap in 16.5~ms. This operation precedes the excitation into the $p$-band of the $A$ triangular lattice by 5~ms. The final intensity of the optical dipole trap is optimized according to the effect of the evaporative cooling. The very shallow trap along the gravity direction provides a good channel for the thermal atoms to escape from the trap without affecting the coherence of the atomic cloud.

\begin{figure}[h]
	\centerline{
	\includegraphics[width=18.3cm]{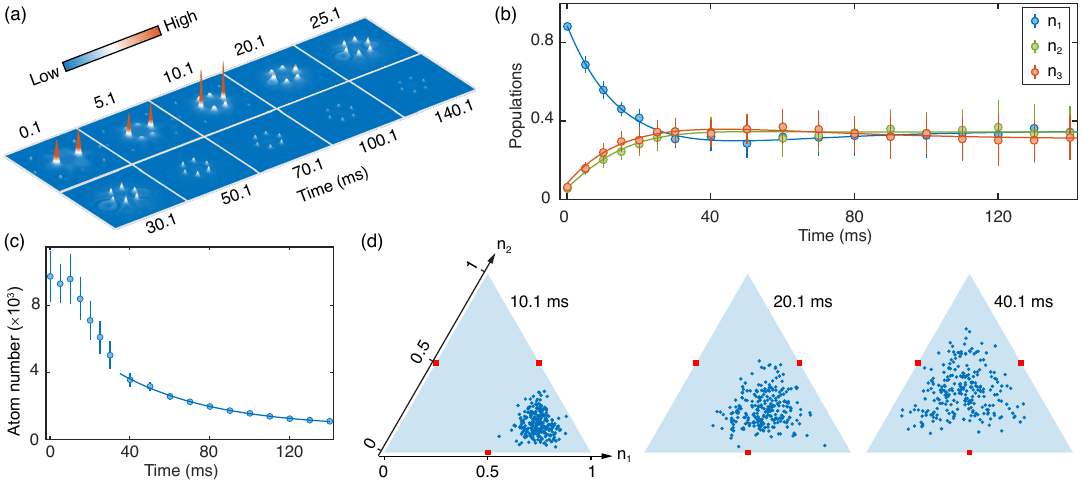}}
	\caption{(a) The evolution of momentum distributions averaged over 51 single-run experimental images. (b) The evolution of the relative populations averaged over 51 experimental runs. The solid lines are fits with exponentially damped harmonic oscillations. Error bars denote standard deviation. (c) The evolution of the total number of atoms populating the $M$-points. The solid line shows an exponential fit for the data after 40 ms. Error bars denote standard deviations. (d) Statistical distributions of the states at early evolution times, each showing 249 experimental runs indicated by blue dots. The ground states predicted by theory are indicated by red squares.}
	\label{sfigex1}
\end{figure}

We observe the evolution of the atoms in the $p$-band of the $A$-triangular lattice by varying the holding time in the $p$-band. The momentum  spectrum of the atoms is recorded by suddenly switching off all potentials followed by a free expansion of 25 ms before an absorption image is taken. For example, Fig.~\ref{sfigex1} shows the condensation dynamics in the maximally symmetric triangular lattice. The evolution of momentum distributions averaged over 51 single-run experimental images are shown in Fig.~\ref{sfigex1}(a). The corresponding evolution of relative populations for atoms residing at three M points are shown in Fig.~\ref{sfigex1}(b). Fig.~\ref{sfigex1}(c) illustrates the evolution of the total number of atoms populating at three M points.  For the maximally symmetric case, in a relatively short time interval of about 40 ms, the system quickly thermalizes to states with approximately equal populations at three M points. This is demonstrated in Fig.~\ref{sfigex1}(d), where statistical distributions of the states at three early evolution times are shown.

\section{\bf Band structure}
\begin{figure}[htbp]
	\centering
	\includegraphics[width=12cm]{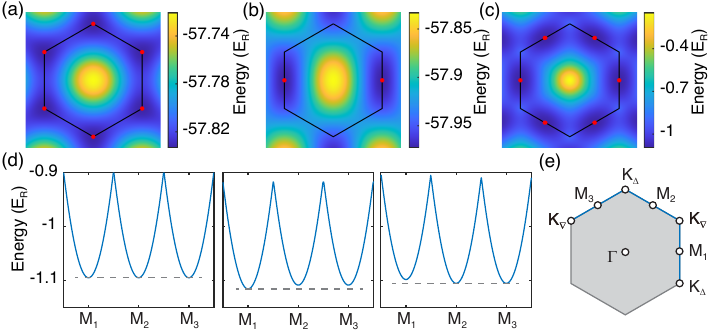}
	\caption{(a, b, c) The energy surfaces of the second band with (a) $(V_A,V_B)=(7.82,7.20)\ E_\mathrm{R}$ and $\Delta\nu = 3.308$ GHz, (b) $(V_A,V_B)=(7.82,7.20)\ E_\mathrm{R}$ and $\Delta\nu = 3.25$ GHz, (c) $(V_A,V_B)=(0.71,0)\ E_\mathrm{R}$. The red dots indicate the positions of the energy minima. (d) The energy distributions of the second band along the high-symmetry lines shown in (e). From left to right $\boldsymbol{\alpha}=(1, 1, 1)$, $(1, 1.01, 1.01)$, and $(1.01, 1, 1)$, respectively. Here, we also choose $V_A=0.71\,E_R$. The gray dashed lines denote the energy minima of the second band.}
	\label{sfig2}
\end{figure}

We calculate exact band structures by the method of plane wave expansion. The hexagonal boron nitride lattice, described by the potential of Eq.~(\ref{bn}), arises for $\Delta\nu=3.308$ GHz, with energy minima of the second band located at the $K$ points. When $\Delta\nu=3.25$ GHz, the optical lattice becomes a deformed hexagonal lattice, whose energy minimum of the second band  is located at the $M_1$ point (see Fig.~\ref{sfig2} (b)). We may hence use the deformed hexagonal lattice to transfer atoms into the $M_1$ point of the second band. This is a convenient starting point for loading atoms into the second band of the triangular lattice whose energy minima are also located at the $M$ points.

Next, we discuss the triangular lattice described by the potential of Eq.~(\ref{potential}). When $\boldsymbol{\alpha}=(1, 1, 1)$, the second band has three degenerate single-particle energy minima located at the $M$ points. Due to the interaction effect, atoms prefer to condense at two out of three $M$ points. To simplify the symmetry breaking process, we can adjust the relative intensity of the three laser beams, forming the lattice potential, in order to lift the energy of one $M$ point. In Fig.~\ref{sfig2} (d), we show two types of energy spectra with $E_2(M_1)<E_2(M_2)=E_2(M_3)$ for $\boldsymbol{\alpha}=(1, 1.01, 1.01)$ and $E_2(M_2)=E_2(M_3)<E_2(M_1)$ for $\boldsymbol{\alpha}=(1.01, 1, 1)$, where $E_n(M_j)$ indicates the single-particle energy at the $M_j$ point for the $n$-th band.

\section{\bf Two-band model}
In this section, we only focus on the second and third bands corresponding to two degenerate $p_x$ and $p_y$ orbitals. We construct the two-band model based on the Bloch functions and Wannier functions, respectively. The ground state can be obtained by the mean-field approximation.

\subsection{Plane wave expansion}
\label{Plane}
The single-particle energy spectra and Bloch functions can be calculated by the plane wave expansion. We only consider six Bloch functions as the basis functions, which are located at the three $M$ points of the second and third bands. Their positions in reciprocal space are $\mathbf{M}_1=k_L(\sqrt{3}/2,0)$, $\mathbf{M}_2=k_L(\sqrt{3}/4,3/4)$, and $\mathbf{M}_3=k_L(-\sqrt{3}/4,3/4)$. The ansatz of the ground state takes the form $\Psi(\mathbf{r})=\sqrt{N}\sum_{n,j}z_{n,j}\Phi_{n,\mathbf{M}_j}(\mathbf{r})$, where $\Phi_{n,\mathbf{M}_j}(\mathbf{r})$ are real Bloch functions with $j=1,2,3$ and $n=2,3$ denote the band index. $N$ is the total particle number and $z_{n,j}$ are six complex numbers with the normalization condition $\sum_{n,j}|z_{n,j}|^2=1$.

Then the total mean-field energy has the form
\begin{eqnarray}
E[\mathbf{z}]=N\sum_{n,j}|z_{n,j}|^2E_n(M_j) +\frac{g}{2}\int |\Psi(\mathbf{r})|^4 d\mathbf{r}.
\end{eqnarray}
Here, $g=4\pi\hbar^2a_s/m$ is positive for the repulsive interaction with $a_s$ denoting the $s$-wave scattering length. The final ground state is obtained by the numerical optimization of $E[\mathbf{z}]$ with a fixed $N$. We find that it has only two quasi-momentum components and their phase difference is $\pm \pi/2$. A general form of the ground state reads
\begin{eqnarray}
\Psi(\mathbf{r})=\sqrt{N}\left\{\cos\xi\left[\Phi_{2,\mathbf{M}_i}(\mathbf{r})\pm i\Phi_{2,\mathbf{M}_j}(\mathbf{r})\right]+\sin\xi\left[\Phi_{3,\mathbf{M}_i}(\mathbf{r})\pm i\Phi_{3,\mathbf{M}_j}(\mathbf{r})\right]\right\}/\sqrt{2}.
\end{eqnarray}
Here, $\xi$ depends on parameters and is determined numerically. When the interaction is weak, $|\sin\xi|\ll 1$, which indicates that most atoms occupy the second band.

\subsection{Tight-binding approximation}
\begin{figure}[htbp]
	\centering
	\includegraphics[width=12cm]{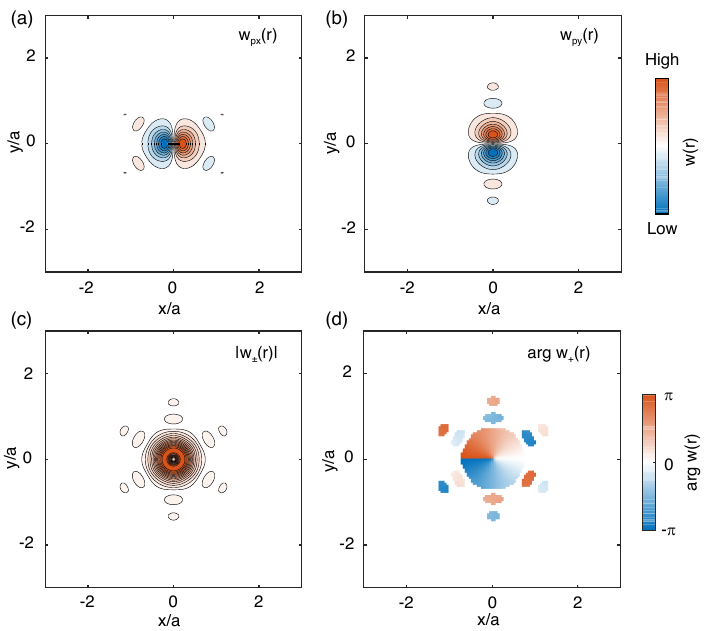}
	\caption{(a) The Wannier function of $p_x$ orbital. (b) The Wannier function of $p_y$ orbital. (c) The modulus of wannier functions $w_{\pm}(\mathbf{r})$. (d) The phase distribution of Wannier function $w_{+}(\mathbf{r})$.}
	\label{sfig3}
\end{figure}

Another scheme to construct a theoretical model of the second and third bands is using the bases of Wannier functions, which are obtained by the Marzari-Vanderbilt method numerically. The shapes of $p_x$ and $p_y$ Wannier functions are slightly different and their superpositions $w_{\pm}(\mathbf{r})=[w_{p_x}(\mathbf{r})+iw_{p_y}(\mathbf{r})]/\sqrt{2}$ satisfy the six-fold rotational symmetry, which is demonstrated in Fig.~\ref{sfig3}.

\begin{figure}[htbp]
	\centering
	\includegraphics[width=12cm]{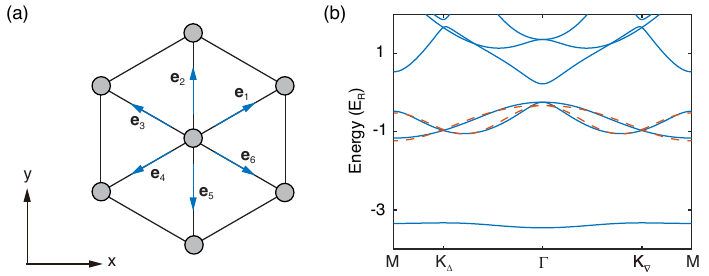}
	\caption{(a) Sketch of the triangular lattice. (b) The comparison between the exact and the tight-binding energy bands. The solid blue lines indicate the band structures from the plane wave expansion with $V_A=0.73E_R$. The orange red dashed lines indicate the tight-binding energy spectra with $t_1=0.069E_R (\sim 6.6859\, \rm nK)$  and $t_2=0.089E_R (\sim8.6517\, \rm nK)$. }
	\label{sfig4}
\end{figure}

Only considering the nearest-neighbor hoppings in a triangular lattice, the single-particle Hamiltonian in the bases of $w_{\pm}(\mathbf{r})$ reads
\begin{eqnarray}
\hat{H}_0=\sum_{\mathbf{R},l}t_1\left(\hat{p}_{+,\mathbf{R}}^\dagger\hat{p}_{+,\mathbf{R}+\mathbf{e}_l}+\hat{p}_{-,\mathbf{R}}^\dagger\hat{p}_{-,\mathbf{R}+\mathbf{e}_l}\right)
+\sum_{\mathbf{R},l}\left(t_2e^{i\nu_l}\hat{p}_{+,\mathbf{R}}^\dagger\hat{p}_{-,\mathbf{R}+\mathbf{e}_l}+\mathrm{h.c.}\right).
\end{eqnarray}
Here, the hopping coefficients satisfy $0<t_1<t_2$ and $\nu_l=(-2l+1)\pi/3$ with $l=1,\cdots,6$. The vectors connecting nearest-neighbor sites are $\mathbf{e}_1=a(\sqrt{3}/2,1/2)$, $\mathbf{e}_2=a(0,1)$, $\mathbf{e}_3=a(-\sqrt{3}/2,1/2)$, $\mathbf{e}_4=-\mathbf{e}_1$, $\mathbf{e}_5=-\mathbf{e}_2$, and $\mathbf{e}_6=-\mathbf{e}_3$ with the lattice constant $a=4\pi/(3k_L)$. After the Fourier transformation, the single-particle Hamiltonian in quasi-momentum space can be written as $\hat{H}_0=\sum_\mathbf{k}(\hat{p}_{+,\mathbf{k}}^\dagger,\hat{p}_{-,\mathbf{k}}^\dagger)\mathcal{H}_0(\mathbf{k})(\hat{p}_{+,\mathbf{k}},\hat{p}_{-,\mathbf{k}})^\mathrm{T}$ with
\begin{eqnarray}
\mathcal{H}_0\left( \mathbf{k} \right) =2\left( \begin{matrix}
	t_1\sum_{l=1}^3{\cos}\left( \mathbf{k}\cdot \mathbf{e}_l \right)&		t_2\left[ e^ { -i\pi/3 } \cos \left( \mathbf{k}\cdot \mathbf{e}_1 \right) -\cos \left( \mathbf{k}\cdot \mathbf{e}_2 \right) +e^ { i\pi/3 } \cos \left( \mathbf{k}\cdot \mathbf{e}_3 \right) \right]\\
	\mathrm{h.c.}&		t_1\sum_{l=1}^3{\cos}\left( \mathbf{k}\cdot \mathbf{e}_l \right)
\end{matrix} \right).
\end{eqnarray}
The tight-binding energy spectrum is shown in Fig.~\ref{sfig4} (b).

The on-site interaction Hamiltonian reads
\begin{eqnarray}
\hat{H}_I=\frac{U_{++}}{2}\sum_\mathbf{R} \left(\hat{p}_{+,\mathbf{R}}^\dagger\hat{p}_{+,\mathbf{R}}^\dagger\hat{p}_{+,\mathbf{R}}\hat{p}_{+,\mathbf{R}}+\hat{p}_{-,\mathbf{R}}^\dagger\hat{p}_{-,\mathbf{R}}^\dagger\hat{p}_{-,\mathbf{R}}\hat{p}_{-,\mathbf{R}}+4\hat{p}_{+,\mathbf{R}}^\dagger\hat{p}_{-,\mathbf{R}}^\dagger\hat{p}_{-,\mathbf{R}}\hat{p}_{+,\mathbf{R}}\right).
\end{eqnarray}
Here, the interaction coefficient $U_{++}$ is defined as $U_{++}=g\int|w_{+}(\mathbf{r})|^4d\mathbf{r}$.

We can also use the bases of $p_{x,y}$ orbitals via applying the transformation $\hat{p}_{\pm}^\dagger=(\hat{p}_{x}^\dagger\pm\hat{p}_{y}^\dagger)/\sqrt{2}$. The single-particle Hamiltonian reads $\hat{H}_0=\sum_\mathbf{k}(\hat{p}_{x,\mathbf{k}}^\dagger,\hat{p}_{y,\mathbf{k}}^\dagger)\left[\mathcal{H}_\sigma(\mathbf{k})+\mathcal{H}_\pi(\mathbf{k})\right](\hat{p}_{x,\mathbf{k}},\hat{p}_{y,\mathbf{k}})^\mathrm{T}$ with
\begin{eqnarray}
\mathcal{H}_\sigma\left( \mathbf{k} \right) =t_{\sigma}\left( \begin{matrix}
\frac{3}{2}\cos \left( \mathbf{k}\cdot \mathbf{e}_1 \right) +\frac{3}{2}\cos \left( \mathbf{k}\cdot \mathbf{e}_3 \right) &		 \frac{\sqrt{3}}{2}\cos \left( \mathbf{k}\cdot \mathbf{e}_1 \right) -\frac{\sqrt{3}}{2}\cos \left( \mathbf{k}\cdot \mathbf{e}_3 \right) \\
\mathrm{h.c.}&		\frac{1}{2}\cos \left( \mathbf{k}\cdot \mathbf{e}_1 \right) +2\cos \left( \mathbf{k}\cdot \mathbf{e}_2 \right) +\frac{1}{2}\cos \left( \mathbf{k}\cdot \mathbf{e}_3 \right)\\
\end{matrix} \right)
\end{eqnarray}
and
\begin{eqnarray}
\mathcal{H}_\pi\left( \mathbf{k} \right) =t_{\pi}\left( \begin{matrix}
\frac{1}{2}\cos \left( \mathbf{k}\cdot \mathbf{e}_1 \right) +2\cos \left( \mathbf{k}\cdot \mathbf{e}_2 \right) +\frac{1}{2}\cos \left( \mathbf{k}\cdot \mathbf{e}_3 \right) &		 -\frac{\sqrt{3}}{2}\cos \left( \mathbf{k}\cdot \mathbf{e}_1 \right) +\frac{\sqrt{3}}{2}\cos \left( \mathbf{k}\cdot \mathbf{e}_3 \right)\\
\mathrm{h.c.}&		\frac{3}{2}\cos \left( \mathbf{k}\cdot \mathbf{e}_1 \right) +\frac{3}{2}\cos \left( \mathbf{k}\cdot \mathbf{e}_3 \right)\\
\end{matrix} \right).
\end{eqnarray}
Here, $t_\sigma>0$ and $t_\pi<0$ are hopping coefficients corresponding to $\sigma$- and $\pi$-type bondings of $p$ orbitals, respectively. The interaction Hamiltonian becomes
\begin{equation}
\begin{aligned}
\hat{H}_I=&\frac{U_{xx}}{2}\sum_\mathbf{R} \left(\hat{p}_{x,\mathbf{R}}^\dagger\hat{p}_{x,\mathbf{R}}^\dagger\hat{p}_{x,\mathbf{R}}\hat{p}_{x,\mathbf{R}}+\hat{p}_{y,\mathbf{R}}^\dagger\hat{p}_{y,\mathbf{R}}^\dagger\hat{p}_{y,\mathbf{R}}\hat{p}_{y,\mathbf{R}}\right)\\
&+\frac{U_{xy}}{2}\sum_\mathbf{R} \left(\hat{p}_{x,\mathbf{R}}^\dagger\hat{p}_{x,\mathbf{R}}^\dagger\hat{p}_{y,\mathbf{R}}\hat{p}_{y,\mathbf{R}}+\hat{p}_{y,\mathbf{R}}^\dagger\hat{p}_{y,\mathbf{R}}^\dagger\hat{p}_{x,\mathbf{R}}\hat{p}_{x,\mathbf{R}}+4\hat{p}_{x,\mathbf{R}}^\dagger\hat{p}_{y,\mathbf{R}}^\dagger\hat{p}_{y,\mathbf{R}}\hat{p}_{x,\mathbf{R}}\right).
\end{aligned}
\end{equation}
Here, the interaction coefficients are $U_{xx}=g\int w_{p_x}^4(\mathbf{r})d\mathbf{r}$ and $U_{xy}=g\int w_{p_x}^2(\mathbf{r})w_{p_y}^2(\mathbf{r})d\mathbf{r}$. Compare with the case of $p_\pm$ orbitals, we obtain $t_\sigma=t_1+t_2$, $t_\pi=t_1-t_2$, $U_{++}=2U_{xx}/3$, and $U_{xx}=3U_{xy}$.

We then only focus on three $M$ points and apply the mean-field approximation to obtain
\begin{eqnarray}
\langle \hat{p}_{\alpha,\mathbf{R}} \rangle=\sqrt{\rho}\left\{ e^{i\mathbf{M}_1\cdot\mathbf{R}}z_{\alpha,1}+e^{i\mathbf{M}_2\cdot\mathbf{R}}z_{\alpha,2}+e^{i\mathbf{M}_3\cdot\mathbf{R}}z_{\alpha,3}\right\},
\end{eqnarray}
where $\alpha=+,-$ or $x,y$ and $\rho=N/N_u$ is the particle number per unit cell with $N_u$ being the number of unit cells. The ground state is obtained by minimizing the mean-field energy under the constraint $\sum_{\alpha,j} |z_{\alpha,j}|^2=1$. We find that the final ground state only has two quasi-momentum components and their phase difference is fixed at $\pm \pi/2$. Suppose that the ground state condenses at $M_1$ and $M_2$ points and the phase difference takes $\pi/2$, the ground state in the bases of $p_{x,y}$ can be written as
\begin{eqnarray}
	\Psi(\mathbf{R})=\sqrt{\rho}\left\{e^{i\mathbf{M}_1\cdot\mathbf{R}}
\left(
\begin{matrix}
	\cos\varphi\\
	-\sin\varphi
\end{matrix}
\right)
+
i e^{i\mathbf{M}_2\cdot\mathbf{R}}
\left(
\begin{matrix}
	\cos(\frac{\pi}{3}+\varphi)\\
	\sin(\frac{\pi}{3}+\varphi)
\end{matrix}
\right)\right\}/\sqrt{2}.
\label{gs}
\end{eqnarray}
Here, the angle $\varphi$ increases with the interaction strength and is equal to $0$ and $\pi/12$ for $U_{xx}\to 0$ and $U_{xx}\to\infty$, respectively. We can also write the ground state in the bases of $p_{\pm}$ by the transformation $\hat{p}_{\pm}^\dagger=(\hat{p}_{x}^\dagger\pm\hat{p}_{y}^\dagger)/\sqrt{2}$.

\section{\bf One-band model}
\label{One}
In the weak interaction limit, atoms mainly condense at the second band. We thus can construct a low-energy one-band model, only focusing on the second band, which captures the major physics of the ground state. In the basis of Bloch functions, the Hamiltonian can be written as
\begin{equation}
\begin{aligned}
\hat{H}=&\sum_{j=1,2,3}E_2(M_j)\hat{b}_{j}^\dagger\hat{b}_{j}+\frac{U_1}{2}\sum_{j=1,2,3}\hat{b}_{j}^\dagger\hat{b}_{j}^\dagger\hat{b}_{j}\hat{b}_{j}\\
&+\frac{U_2}{2}\sum_{j=1,2,3}\left(\hat{b}_{j}^\dagger\hat{b}_{j}^\dagger\hat{b}_{j+1}\hat{b}_{j+1}+\hat{b}_{j}^\dagger\hat{b}_{j}^\dagger\hat{b}_{j+2}\hat{b}_{j+2}+2\hat{b}_{j}^\dagger\hat{b}_{j+1}^\dagger\hat{b}_{j+1}\hat{b}_{j}+2\hat{b}_{j}^\dagger\hat{b}_{j+2}^\dagger\hat{b}_{j+2}\hat{b}_{j}\right).
\label{OneBand}
\end{aligned}
\end{equation}
Here, $\hat{b}_{j}$ ($\hat{b}_{j}^\dagger$) indicates annihilating (creating) a boson with the Bloch state $\Phi_{2,\mathbf{M}_j}$ at $M_j$ point. $j=1,2,3$ and the addition in the $j$ index is modulo 3. $U_{1,2}$ are defined as $U_1=g\int |\Phi_{2,\mathbf{M}_1}(\mathbf{r})|^4d\mathbf{r}$ and $U_2=g\int \Phi_{2,\mathbf{M}_1}^{*2}(\mathbf{r})\Phi_{2,\mathbf{M}_2}^2(\mathbf{r})d\mathbf{r}$. Note that the Bloch functions $\Phi_{2,\mathbf{M}_j}(\mathbf{r})$ is real, so that $U_{2}$ is also real.

We roughly estimate that the number of atoms per unit cell is about 10, which closely matches the real experimental condition after a long time relaxation, e.g. 140.1 ms. Via numerical calculation, we find $NU_1=0.0431E_\mathrm{R}$ ($\sim 4.1974\mathrm{nK}$) and $NU_2=0.0236E_\mathrm{R}$ ($\sim 2.2959\mathrm{nK}$) when $V_A=0.73E_\mathrm{R}$.

Based on the mean-field approximation, we replace the operators $\hat{b}_{j}$ by $\sqrt{N}z_j$ with the normalization condition $\sum_j|z_j|^2=1$. Then the mean-field energy can be written as $E[\mathbf{z}]=N\sum_j|z_j|^2E_2(M_j)+E_{\rm int}[\mathbf{z}]$, where the interaction energy is given by 
\begin{eqnarray}
E_{\rm int}[\mathbf{z}]=\sum_{j}\left\{\frac{U_1N^2}{2}|z_j|^4+\frac{U_2N^2}{2}\left[(z_j^*)^2(z_{j+1})^2+(z_j^*)^2(z_{j+2})^2+2|z_j|^2|z_{j+1}|^2+2|z_j|^2|z_{j+2}|^2\right]\right\}.
\label{OneMF}
\end{eqnarray}

For the symmetric case where $E_2(M_1)=E_2(M_2)=E_2(M_3)$, the single-particle part is a constant with varying $z_j$ and can be neglected. Minimizing the mean-field energy, we obtain six degenerate ground states, whose order parameter are given by $\sqrt{N}[\Phi_{2,\mathbf{M}_j}(\mathbf{r})\pm i\Phi_{2,\mathbf{M}_{j+1}}(\mathbf{r})]/\sqrt{2}$. In the ground state, only two of three $M$ points are populated and their phase difference is locked as $\pm \pi/2$. An exemplary degenerate ground state is given by $\mathbf{z}=(z_1,z_2,z_3)=(1,\pm i,0)/\sqrt{2}$.

\begin{figure}[htbp]
	\centering
	\includegraphics[width=12cm]{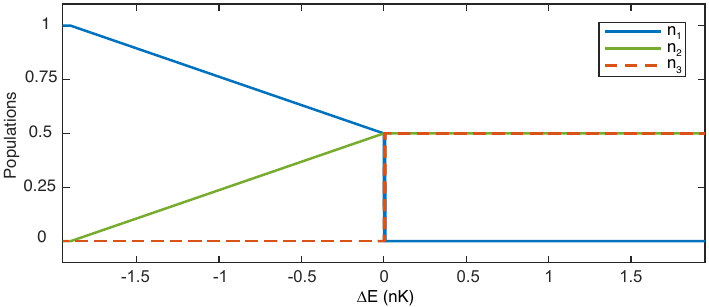}
	\caption{The $M$-point populations of ground state for different energy imbalance $\Delta E$.}
	\label{sfig5}
\end{figure}

We then discuss the ground state when there are energy imbalances among the three $M$ points of the second band, which correspond to the intensity imbalances of the three lattice laser beams in the experiment. To simplify the discussion, we only shift the energy of the $M_1$ point and set the energies of the $M_2$ and $M_3$ points fixed and equal. The energy imbalance has been defined as $\Delta E\equiv E_2(M_1)-[E_2(M2)+E_2(M_3)]/2$ in the main text. We use the one-band model of Eq.~(\ref{OneBand}) to calculate the ground state for different $\Delta E$ at the mean-field level. In the calculation, $NU_1=0.0431E_\mathrm{R}$ and $NU_2=0.0236E_\mathrm{R}$ are fixed. The dependence of the populations at the three $M$ points of the ground state on the energy imbalance $\Delta E$ is shown in Fig.~\ref{sfig5}. Note that here we only consider degenerate ground states with $n_2 > n_3$ and ignore degenerate states with $n_2< n_3$  for the case $\Delta E<0$.

When $\Delta E$ is far smaller than zero, the ground state atoms only condense at the $M_1$ point with $n_1=1$; When $\Delta E$ is a small negative number, the majority of the atoms prefers to condense at $M_1$ and the minority of the atoms condenses at $M_2$ or $M_3$. The relative phase between the condensate at two $M$ points is fixed at $\pm \pi/2$; When $\Delta E>0$, the atoms prefer to condense at the $M_2$ and $M_3$ points with equal populations with $n_2=n_3=0.5$. Their relative phase is also fixed at $\pm \pi/2$. These results are consistent with experimental data shown in Fig.~2(b) of the main text.

\begin{figure}[htbp]
\centering
\includegraphics[width=12cm]{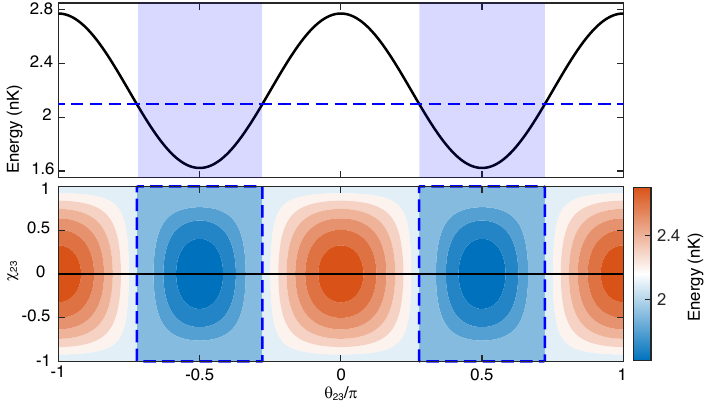}
\caption{ In the lower panel, the mean-field energy per atom is plotted versus the contrast $\chi_{23}$ and the relative phase $\theta_{23}$. The upper panel shows the profile along the black solid horizontal line in the center of the lower panel. The blue dashed lines in the lower panel denotes the contours with an energy given by the blue dashed line in the upper panel. The value of the unknown phase $\theta_{23}$ can be restricted to the blue translucent area of the upper panel by the reasonable assumption that the system tends to minimize its energy during relaxation.}
\label{sfig6}
\end{figure}

Experimentally, the population ratios $n_{1,2,3}$ for atoms at $M_{1,2,3}$ can be directly inferred from any TOF image. However, this is not the case for the phase information of $z_j$. We parameterize $\mathbf{z}$ as $(\sqrt{n_1}\exp(i\theta_1),\sqrt{n_2}\exp(i\theta_2),\sqrt{n_3}\exp(i\theta_3))$ with three phases $\theta_{1,2,3}$ to be determined. We can infer the desired phase information from a simple mean-field consideration upon the assumption that the system tends to minimize its energy. Since the system tends to break the lattice rotational symmetry and condenses at two out of three $M$ points, we consider the example of a superposition of equal portions of the associated second band Bloch functions $\Phi_{2,\mathbf{M}_{2}}(\mathbf{r})$ and $\Phi_{2,\mathbf{M}_{3}}(\mathbf{r})$. The mean-field energy $E[\mathbf{z}]$ versus $\chi_{23}\equiv(n_2-n_3)/(n_2+n_3)$ and the relative phase $\theta_{23}$ between the condensate populations is shown in Fig.~\ref{sfig6}. Note that the blue dashed rectangles denote the contours of the mean-field energy of the initial state, in which all atoms condense at the $M_1$ point. Hence, the mere assumption that the ground state energy must be lower than the initial state, restricts the phase difference among the condensates at the $M_2$ and $M_3$ points to the region centered at $\theta_{23}=\pm\pi/2$ with an uncertainty roughly fixed to $\pm\,\pi/4$. Thus, the time-reversal symmetry preserved states with $\theta_{23}=0$ or $\pi$ are excluded, which suggests that the unconventional Bose-Einstein condensate we observe should in fact break time-reversal symmetry.

\begin{figure}[htbp]
	\centering
	\includegraphics[width=12cm]{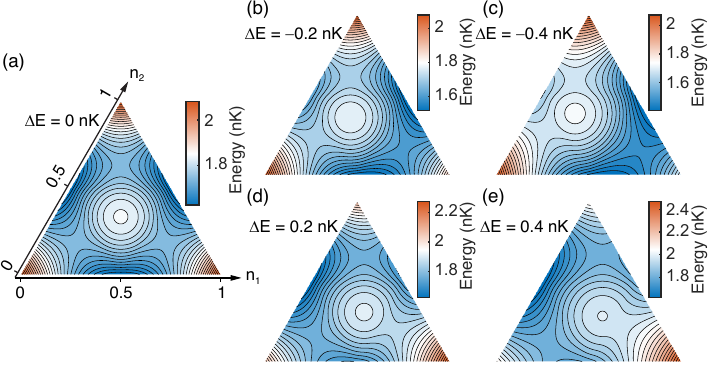}
	\caption{The mean-field energy per atom versus arbitrary $M$-point populations for the energy imbalance (a) $\Delta E = 0\,\mathrm{nK}$, (b) $\Delta E = -0.2\,\mathrm{nK}$, (c) $\Delta E = -0.4\,\mathrm{nK}$, (d) $\Delta E = 0.2\,\mathrm{nK}$, and (e) $\Delta E = 0.4\,\mathrm{nK}$. The phases $\theta_{1,2,3}$ are chosen for the optimal energy.}
	\label{sfigex2}
\end{figure}

For the fixed $M$-point population ratios $n_{1,2,3}$, the optimal phases $\theta_{1,2,3}$ are determined by minimizing the mean-field energy. By choosing the optimal phases, the mean-field energy for the arbitrary population ratios can be obtained. We find that the $M$-point populations are very sensitive for the energy imbalance $\Delta E$ in the experiment, which is consistent with theoretical calculations shown in Fig.~\ref{sfigex2}.

\section{\bf Loop current}
\begin{figure}[htbp]
	\centering
	\includegraphics[width=12cm]{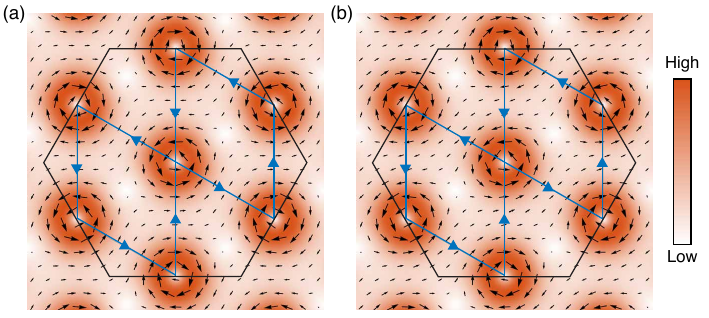}
	\caption{(a) The mass current of the ground state based on the one-band plane-wave model. (b) The mass current of the ground state based on the two-band plane wave model. Here, the amplitude (direction) of the current is shown by a color code (arrows).}
	\label{sfig7}
\end{figure}

The mass current for an arbitrary wave function is defined as
\begin{eqnarray}
\mathbf{j}(\mathbf{r})=\frac{i\hbar}{2m}\left[\Psi(\mathbf{r})\nabla\Psi^*(\mathbf{r})-\Psi^*(\mathbf{r})\nabla\Psi(\mathbf{r})\right].
\end{eqnarray}
Here, the gradient of the wave function can be calculated by the finite difference method. We calculate the mass current of the ground state for different interaction strengths based on the two-band plane-wave model in Sec.~\ref{Plane}. When the interaction is close to zero, the ground state involves the Bloch states for the second band, which is the same as the result from the one-band model in Sec.~\ref{One}. Mass currents for the ground states including and not including the third band components are shown in Fig.~\ref{sfig7}.

We can also derive the mass current based on the tight-binding model. First we define the multiple-orbital polarization operator as
\begin{eqnarray}
\hat{\mathbf{P}}=\sum_{\alpha,\beta,\mathbf{R}} \mathbf{R}_{\alpha,\beta}\hat{p}_{\alpha,\mathbf{R}}^\dagger \hat{p}_{\beta,\mathbf{R}},
\end{eqnarray}
where $\mathbf{R}_{\alpha,\beta}=\int w_{\alpha,\mathbf{R}}^*(\mathbf{r}) \mathbf{r} w_{\beta,\mathbf{R}}(\mathbf{r}) d\mathbf{r}$. In the triangular lattice, $\mathbf{R}_{\alpha,\beta}=0$, when $\alpha\neq \beta $ due to the symmetry of the Wannier functions. So the polarization operator becomes
\begin{eqnarray}
\hat{\mathbf{P}}=\sum_{\alpha,\mathbf{R}} \mathbf{R}\hat{p}_{\alpha,\mathbf{R}}^\dagger \hat{p}_{\alpha,\mathbf{R}}.
\end{eqnarray}
The mass current operator can be obtained by using the Heisenberg equation of motion
\begin{equation}
\begin{aligned}
\hat{\mathbf{J}}=&\frac{\partial}{\partial t} \hat{\mathbf{P}}=\frac{i}{\hbar}\left[\hat{H},\hat{\mathbf{P}}\right]\\
=&\frac{i}{\hbar}\sum_{\alpha,\beta}\sum_{\mathbf{R},l}\mathbf{e}_lt_{\alpha,\mathbf{R};\beta,\mathbf{R}+\mathbf{e}_l}\hat{p}_{\alpha,\mathbf{R}}^\dagger \hat{p}_{\beta,\mathbf{R}+\mathbf{e}_l}.
\end{aligned}
\end{equation}
Here, $t_{\alpha,\mathbf{R};\beta,\mathbf{R}+\mathbf{e}_l}$ denotes the hopping coefficient between $w_{\alpha,\mathbf{R}}(\mathbf{r})$ and $w_{\beta,\mathbf{R}+\mathbf{e}_l}(\mathbf{r})$.

\begin{figure}[htbp]
	\centering
	\includegraphics[width=8.9cm]{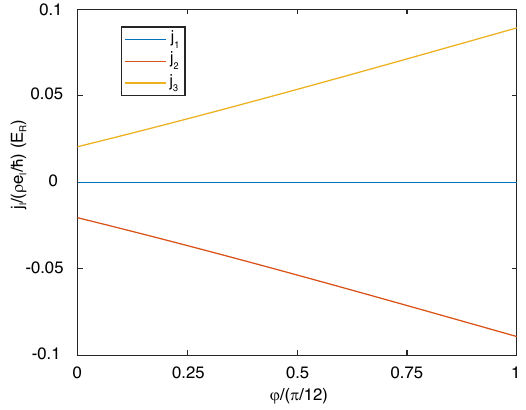}
	\caption{The bond currents $\mathbf{j}_1$, $\mathbf{j}_2$, and $\mathbf{j}_3$ for ground states with $\varphi$.}
	\label{sfig8}
\end{figure}

The bond current operator corresponding to two nearest neighbor sites at $\mathbf{R}$ and $\mathbf{R}+\mathbf{e}_l$ with $\mathbf{R}=\mathbf{0}$ can be written as
\begin{eqnarray}
\hat{\mathbf{j}}_l=\frac{i}{\hbar}\mathbf{e}_l\left(
t_{1} \hat{p}_{+,\mathbf{0}}^\dagger \hat{p}_{+,\mathbf{e}_l}
+t_{1} \hat{p}_{-,\mathbf{0}}^\dagger \hat{p}_{-,\mathbf{e}_l}
+t_2e^{i\nu_l}\hat{p}_{+,\mathbf{0}}^\dagger \hat{p}_{-,\mathbf{e}_l}
+t_2e^{-i\nu_l}\hat{p}_{-,\mathbf{0}}^\dagger \hat{p}_{+,\mathbf{e}_l}
+\mathrm{h.c.}\right).
\end{eqnarray}
Applying the mean-field approximation, we obtain bond currents $\mathbf{j}_l=\langle \hat{\mathbf{j}}_l \rangle_{\Psi}$ for the ground state of Eq.~(\ref{gs}) as
\begin{eqnarray}
\mathbf{j}_1=0, \quad 
\mathbf{j}_2=-j_{\rm bc}\mathbf{e}_2,\quad
\mathbf{j}_3=j_{\rm bc}\mathbf{e}_3,\quad
\mathbf{j}_4=0,\quad
\mathbf{j}_5=-j_{\rm bc}\mathbf{e}_5,\quad
\mathbf{j}_6=j_{\rm bc}\mathbf{e}_6,
\label{bondcurrent}
\end{eqnarray}
where $j_{\rm bc}=\rho[t_2-2t_1\cos(2\varphi+\pi/3)]/\hbar$.
The three bond currents $\mathbf{j}_{1,2,3}$ are shown in Fig.~\ref{sfig8}.

Since experimentally we cannot fully determine the phase difference among two $M$ points, we consider a general state with equal populations at $M_1$ and $M_2$ points with an unknown relative phase $\theta$. The corresponding wave function is given by
\begin{eqnarray}
	\tilde{\Psi}(\mathbf{R})=\sqrt{\rho}\left\{e^{i\mathbf{M}_1\cdot\mathbf{R}}
	\left(
	\begin{matrix}
		\cos\varphi\\
		-\sin\varphi
	\end{matrix}
	\right)
	+
	e^{i\theta} e^{i\mathbf{M}_2\cdot\mathbf{R}}
	\left(
	\begin{matrix}
		\cos(\frac{\pi}{3}+\varphi)\\
		\sin(\frac{\pi}{3}+\varphi)
	\end{matrix}
	\right)\right\}/\sqrt{2}.
\end{eqnarray}
After calculation, we find that the bond currents $\tilde{\mathbf{j}}_l=\langle \hat{\mathbf{j}}_l\rangle_{\tilde{\Psi}}$ for the state $\tilde{\Psi}$ are $\tilde{\mathbf{j}}_l=\mathbf{j}_l\sin\theta$ with $\mathbf{j}_l$ given by Eq.~(\ref{bondcurrent}). When $\theta=\pm \pi/2$, $\tilde{\Psi}$ is the ground state and bond currents take their maximal values. However, nonzero bond currents prevail over the entire regime when $\theta\ne0$ and $\theta\ne\pi$.

Based on the experimental observation and theoretical analysis shown in Fig.~\ref{sfig6}, we exclude the possibility of the phase difference $\theta$ to be close to $0$ or $\pi$. This clearly indicates the existence of loop current order for the experimentally observed unconventional Bose-Einstein condensate where two out of three $M$ points are equally populated.

\section{\bf Domain structure}
\begin{figure}[htbp]
	\centering
	\includegraphics[width=12cm]{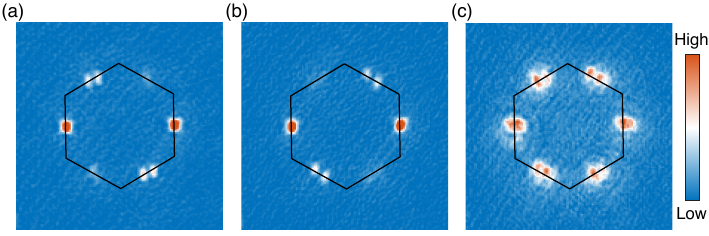}
	\caption{Some experimental TOF images with the domain structures for hold times (a) $10.1$ ms, (b) $10.1$ ms, (c) $15.1$ ms,.}
	\label{sfi8}
\end{figure}

We often observe momentum spectra with Bragg resonances around specific $M$ points split into two sectors, especially after short evolution times (about $5 \sim 20$ ms). Several characteristic TOF images are shown in Fig.~\ref{sfi8}. This experimental phenomenon may originate from the domain formation due to spontaneous symmetry breaking induced by the interaction. In this section, we calculate the momentum distributions of the condensate in presence of domain structures, which agree with the experimental observations.

The momentum distribution of the order parameter is $|\Psi(\mathbf{k})|^2$, where $\Psi(\mathbf{k})$ is obtained by the Fourier transformation
\begin{eqnarray}
\Psi(\mathbf{k})=\frac{1}{\sqrt{\Omega}}\int_{\Omega} e^{-i\mathbf{k}\cdot \mathbf{r}}\Psi(\mathbf{r})d\mathbf{r}
\end{eqnarray}
with $\Omega$ denoting the system volume. Obviously, a Bloch function in momentum space is the summation of delta functions for an infinite integral region. When the integral region is finite, the Fourier transformation can be calculated by the numerical integral method and the delta functions become extended wave packets in momentum space.

\begin{figure}[htbp]
	\centering
	\includegraphics[width=12cm]{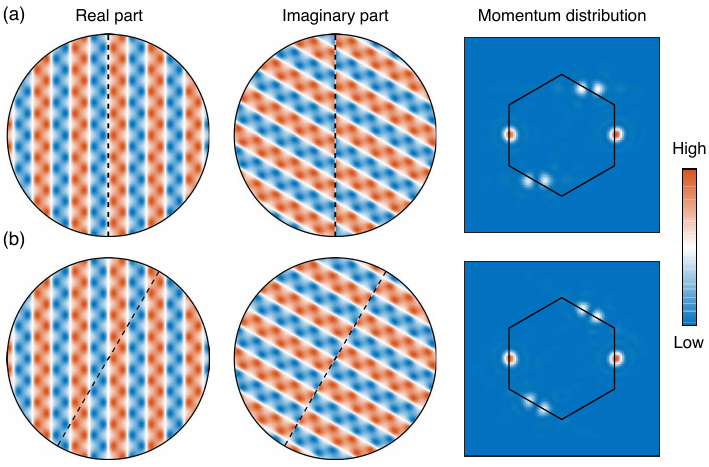}
	\caption{The left and middle pictures show the real and imaginary parts of the order parameter in real space. The dashed black lines indicate the domain walls. The right pictures show the corresponding momentum distributions in reciprocal space. (a) The corresponding order parameter is described by $\sqrt{N}[\Phi_{2,\mathbf{M}_1}(\mathbf{r})-i\Phi_{2,\mathbf{M}_2}(\mathbf{r})]/\sqrt{2}$ for the left part and $\sqrt{N}[\Phi_{2,\mathbf{M}_1}(\mathbf{r})+i\Phi_{2,\mathbf{M}_2}(\mathbf{r})]/\sqrt{2}$ for the right part. (b) is analogous to (a) but with a different orientation of the domain wall.}
	\label{sfig10}
\end{figure}

To approximately describe the real situation, we consider a finite system with a disk geometry.  Assume the order parameter in real space is separated into two sectors with equal area. When the order parameter for one sector takes the from $\sqrt{N}[\Phi_{2,\mathbf{M}_1}(\mathbf{r})-i\Phi_{2,\mathbf{M}_2}(\mathbf{r})]/\sqrt{2}$ and the another sector reads $\sqrt{N}[\Phi_{2,\mathbf{M}_1}(\mathbf{r})+i\Phi_{2,\mathbf{M}_2}(\mathbf{r})]/\sqrt{2}$, we find that the Bragg resonance around the $M_2$ is split into two sectors, as illustrated in Fig.~\ref{sfig10}. In addition, the relative position of the two peaks is influenced by the orientation of the domain wall. These theoretical findings are quite similar to the experimental results. We thus conclude that there is a spontaneous symmetry breaking process as the system approaches equilibrium. For longer evolution time, such kind of domain structure is no longer found in the momentum distribution. We attribute this to the atom loss process in our experiment, which is associated with cooling, with the result that due to interaction only a single domain survives. 

\end{document}